\documentclass{webofc}
\usepackage[varg]{txfonts}   
\begin{document}
\title{A hidden classical symmetry of QCD}
%
%

\author{\firstname{L. Ya. } \lastname{Glozman}\inst{1}\fnsep\thanks{\email{leonid.glozman@uni-graz.at}} 
}

\institute{Institute of Physics, University of Graz, A-8010 Graz, Austria
          }

\abstract{%
  The classical part of the QCD partition function (the integrand) has,
  ignoring irrelevant exact zero modes of the Dirac operator, a local
 $SU(2N_F)  \supset SU(N_F)_L \times SU(N_F)_R \times U(1)_A$ symmetry
 which is absent at the Lagrangian level.
 This symmetry is broken anomalously and spontaneously. Effects of
 spontaneous breaking of chiral symmetry are contained in the near-zero
 modes of the Dirac operator. If physics of anomaly is also encoded in
 the same near-zero modes, then their truncation on the lattice should
 recover a hidden classical $SU(2N_F)$ symmetry in correlators and spectra. This
 naturally explains observation on the lattice of a large degeneracy of hadrons, that is higher than the $SU(N_F)_L \times SU(N_F)_R \times U(1)_A$
chiral symmetry, upon elimination by hands of the lowest-lying modes of the
Dirac operator. We also discuss an implication of this symmetry for the high temperature QCD. 
}
\maketitle
\section{Introduction}
\label{intro}
The QCD Lagrangian in Minkowski space-time has in the chiral limit
the chiral symmetry: 

\begin{equation}
U(N_F)_L \times U(N_F)_R =
SU(N_F)_L \times SU(N_F)_R \times U(1)_A \times U(1)_V.
\label{qcdsymm}
\end{equation}

\noindent
In the following we will always drop the $U(1)_V$ symmetry that is irrelevant
to our subject. The $U(1)_A$ symmetry is  invariance of the Lagrangian upon the axial flavor-neutral transformation

\begin{equation}
\Psi(x) \rightarrow e^{i\alpha \gamma_5}\Psi(x); ~~~~
\bar \Psi(x) \rightarrow \bar \Psi(x) e^{i\alpha \gamma_5}.
\label{u1a}
\end{equation}

\noindent
The $SU(N_F)_L \times SU(N_F)_R$ symmetry implies  invariance
under a pure flavor rotation $SU(N_F)$ ( for $N_F=2$ it is the isospin symmetry)
as well as invariance under the axial flavor transformation

\begin{equation}
\Psi(x) \rightarrow e^{i \gamma_5 \frac{\vec{\lambda}\cdot \vec{\alpha}}
{2}}
\Psi(x); ~~~~
\bar \Psi(x) \rightarrow \bar \Psi(x) e^{i\gamma_5 \frac{\vec{\lambda}\cdot \vec{\alpha}}
{2}},
\label{u2a}
\end{equation}

\noindent
where $\vec{\lambda}$ are $SU(N_F)$ generators.

The axial $U(1)_A$ symmetry is broken anomalously, which is due to
a noninvariance of the integration measure in the functional integral under a local
 $U(1)_A$ transformation  \cite{FU}.
The $ SU(N_F)_A $ "symmetry"\footnote{These transformations do not form a closed
subgroup of the chiral group.} (\ref{u2a}) is broken spontaneously,
because the ground state of the theory, the vacuum, is not invariant under
the transformation (\ref{u2a}). The latter noninvariance is encoded in
the nonzero quark condensate of the vacuum, $<0| \bar\Psi(x) \Psi(x)|0 > \neq 0$.
The quark condensate of the vacuum can be expressed through a density of the
near-zero modes $\lambda \rightarrow 0$ of the Euclidean Dirac operator
(the Banks-Casher relation \cite{BC})

\begin{equation}
\lim_{m \rightarrow 0} <0|\bar \Psi(x) \Psi(x)|0> = -\pi \rho(0)~.
\label{bc}
\end{equation}

The idea of the $N_F=2$ lattice studies \cite{l1,l2} was to understand
what would happen with $J=1,0$ mesons upon truncation of the near-zero modes
of the Dirac operator.
One expects that after truncation correlators of operators that are connected with each other through the $SU(2)_A$
transformation would become identical. If hadrons survive this  truncation, then
masses of chiral partners should  be equal.
It has turned out that a very clean exponential decay of  correlators was detected in all  $J=1$ mesons. This
implies that confined bound  states survive the truncation.
  In the $J=0$
mesons, while all $J=0$ correlators become identical, the ground
states  disappear , because
there is no exponential decay of the corresponding correlators: The near-zero modes  are crucially 
important for  existence of the (pseudo) Goldstone bosons, which is not surprising.

It has also turned out that the truncation restores in hadrons
not only the $SU(2)_L \times SU(2)_R$ symmetry but also the $U(1)_A$ symmetry. One then concludes that the
same lowest-lying modes  are responsible for both
$SU(2)_L \times SU(2)_R$ and $U(1)_A$ breakings which is consistent with the
instanton-induced mechanism of both breakings \cite{H,S,DP,SS}.

However, a larger
degeneracy that includes all possible chiral multiplets of the $J=1$
mesons was detected, which was completely unexpected. This  surprising degeneracy implies a symmetry that is higher than the
$SU(2)_L \times SU(2)_R \times U(1)_A$.
This not yet known symmetry has been reconstructed in refs. \cite{g,gp}
and turned out to be

\begin{equation}
SU(2N_F)  \supset SU(N_F)_L \times SU(N_F)_R \times U(1)_A.
\label{su2nf}
\end{equation}

\noindent
Transformations of this group include both the flavor rotations of the left-
and right-handed quarks as well as transformations that mix the left-
and right-handed components. This symmetry has been confirmed in lattice simulations with the $J=2$
mesons \cite{l3} and in baryons \cite{l4}.

While such an enlarged symmetry, that is larger than the symmetry of the
QCD Lagrangian, was observed in the lattice experiment, its origin was
mysterious. The light on this issue has been shed in ref. \cite{gg}.
It has been shown that the classical part of the QCD partition function has,
excluding irrelevant exact zero modes, a $SU(2N_F)$ local symmetry. This
symmetry is not a symmetry of the Euclidean QCD Lagrangian because the
irrelevant exact zero modes of the Euclidean Dirac operator break it. Consequently, we
refer it as a hidden classical symmetry. This hidden classical symmetry
is broken by the anomaly to $SU(N_F)_L \times SU(N_F)_R$. Spontaneous breaking
of chiral symmetry reduces it to $SU(N_F)_V$. Effects of spontaneous and
anomalous breakings is encoded in the near-zero modes of the Dirac operator.
Consequently, truncation on the lattice of the lowest-lying modes restores
the hidden $SU(2N_F)$ symmetry, which naturally explains  degeneracies observed in lattice experiment.

 Now we will present some details.

\section{What can be apriori expected from the low-mode truncation?}

A truncation of the low-lying modes of the Dirac operator means the following.
The Euclidean Lagrangian  with $N_F$ degenerate  quarks
in a given gauge background is:

\begin{equation}
{\cal {L}} = \Psi^\dag(x)( \gamma_\mu D_\mu + m) \Psi(x),
\label{lag}
\end{equation}

\noindent
where

\begin{equation}
D_\mu = \partial_\mu + i g\frac{t^a}{2} A^a_\mu.
\end{equation}

\noindent
 The hermitian Dirac operator, $i \gamma_\mu D_\mu$,   has in a  volume $V$ a discrete spectrum with real eigenvalues $\lambda_n$:

\begin{equation}
i \gamma_\mu D_\mu  \Psi_n(x) = \lambda_n \Psi_n(x).
\label{ev}
\end{equation}

\noindent
We subtract from the full quark propagator $S_{Full}$ the lowest $k$ eigenmodes
of the Dirac operator

\begin{equation}
S(x,y) =S_{Full}(x,y)-
  \sum_{n=1}^{k}\,\frac{1}{\lambda_n + im}\, \Psi_n (x) \Psi_n^\dagger(y).
\label{prop}
\end{equation}

\noindent
At the same time we keep the gauge configurations intact. Given
these truncated quark propagators we apply standard procedures to extract
hadron spectra using the variational approach. Then we study dependence
of hadron masses on the truncation number $k$. We perform $N_F =2$ dynamical lattice calculations with the overlap Dirac operator with the gauge
configurations generated by the JLQCD collaboration, for  details
see refs. \cite{l1,l2,l3,l4}.

When the near-zero modes, that are responsible for spontaneous chiral symmetry
breaking are subtracted we can expect the $SU(2)_L \times SU(2)_R$ symmetry
in correlators. If in addition the hadron states survive this surgery,
then we can expect a mass degeneracy of chiral partners. For the $J=1$ mesons
the chiral partners are linked by the red arrows on Fig. 1.

\begin{figure}
\centering
\includegraphics[angle=0,width=0.7\linewidth]{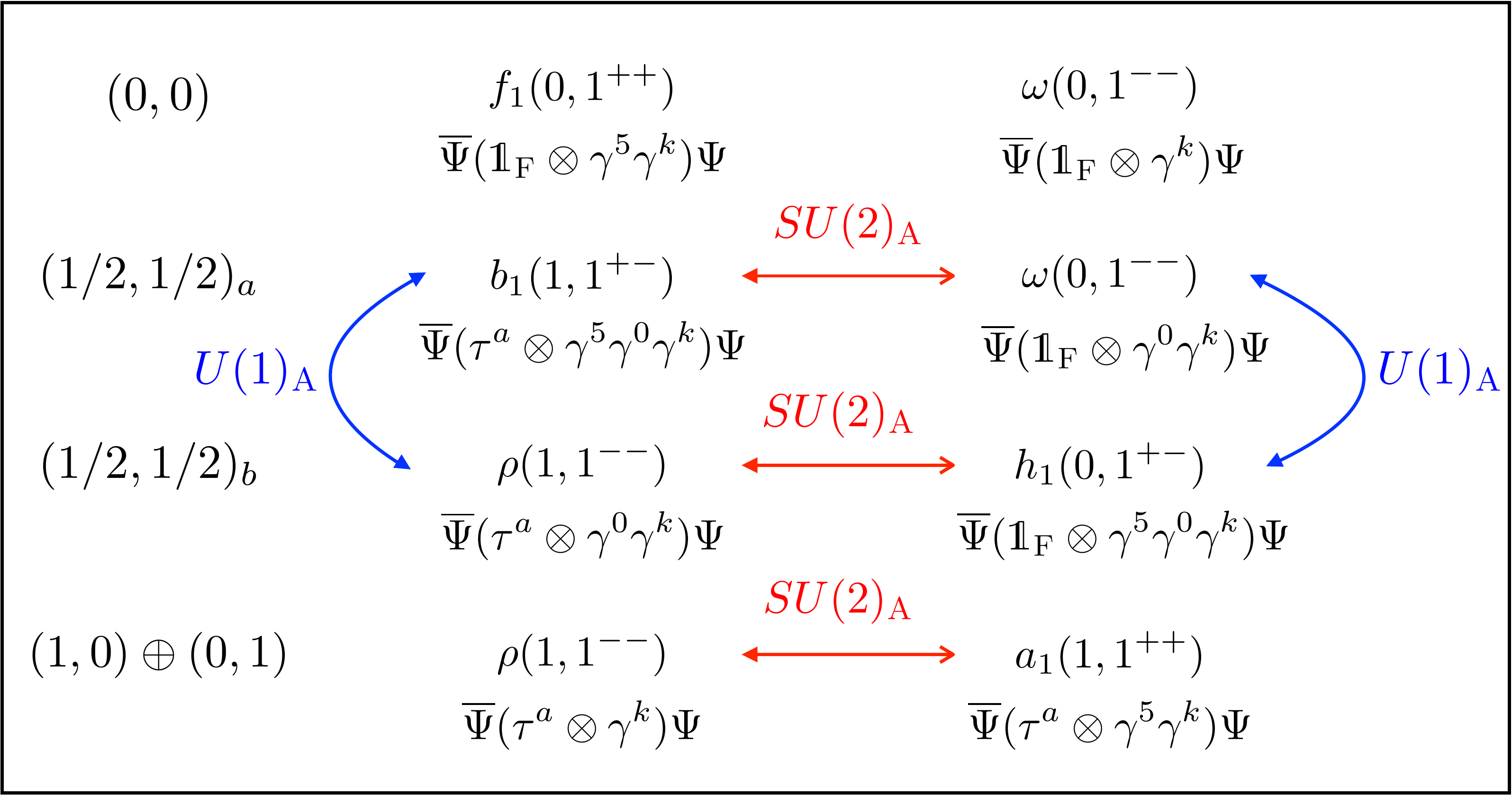}
\caption{$SU(2)_L \times SU(2)_R$ and $U(1)_A$ classification of the $J=1$
meson operators.}
\label{fig-1}
\end{figure}

The $U(1)_A$ transformation connects operators that are linked by the
blue arrows on Fig. 1. Consequently, if the whole chiral symmetry of QCD
$SU(2)_L \times SU(2)_R \times U(1)_A$ is restored, then we can expect
a degeneracy of four mesons from the $(1/2,1/2)_a$ and $(1/2,1/2)_b$
representations on the one hand, and on the other hand a degeneracy
of the $\rho$ and $a_1$ mesons from the $(1,0)+(0,1)$ chiral representation.
Note that in the chirally symmetric world there two independent orthogonal
$\rho$-mesons that belong to two different chiral representations. In the real
world with chiral symmetry breaking these two chiral representations are
mixed in the meson wave function and two different $\rho$ operators couple to one and the same $\rho$-meson. 

The $SU(2)_L \times SU(2)_R \times U(1)_A$ symmetry does not connect,
however, the four mesons from the $(1/2,1/2)_a$ and $(1/2,1/2)_b$
representations with the other mesons in Fig. 1. Consequently, given
only the $SU(2)_L \times SU(2)_R \times U(1)_A$ chiral symmetry we
cannot expect a degeneracy larger than it is shown by arrows on Fig. 1.

\section{Results}

We do not show here our results for the real world ($k=0$), since they
are typical for all lattice studies and the experimental meson spectra
are reasonably reproduced.
The results for the eigenvalues of the cross-correlation matrix
and effective mass plateau at $k=10$ (10 lowest Dirac eigenmodes
have been removed) for the isovector mesons are shown in Fig. 2.

\begin{figure}
\centering 
\includegraphics[width=0.43\textwidth]{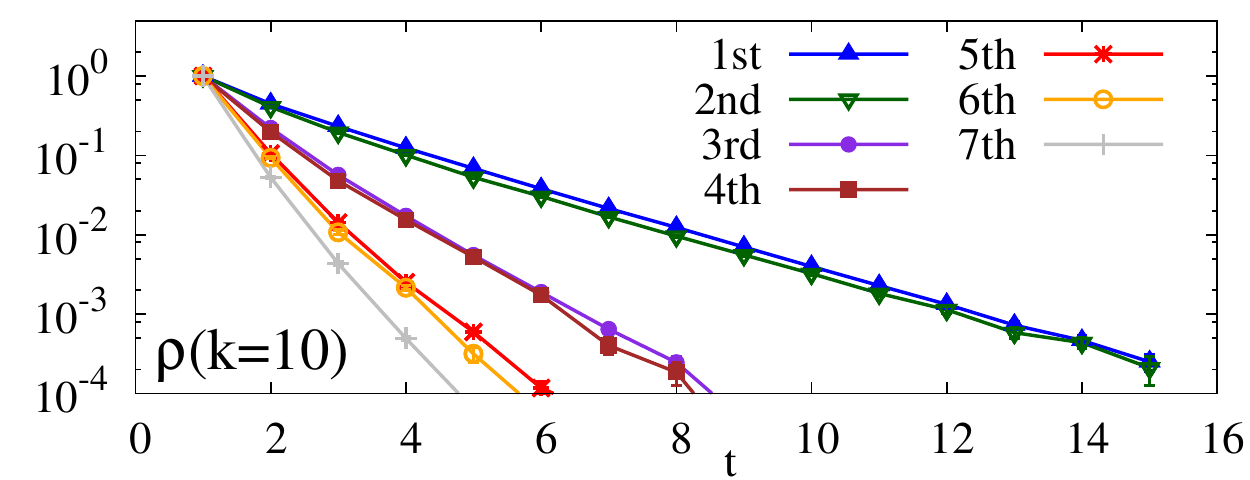}\hfill
\includegraphics[width=0.43\textwidth]{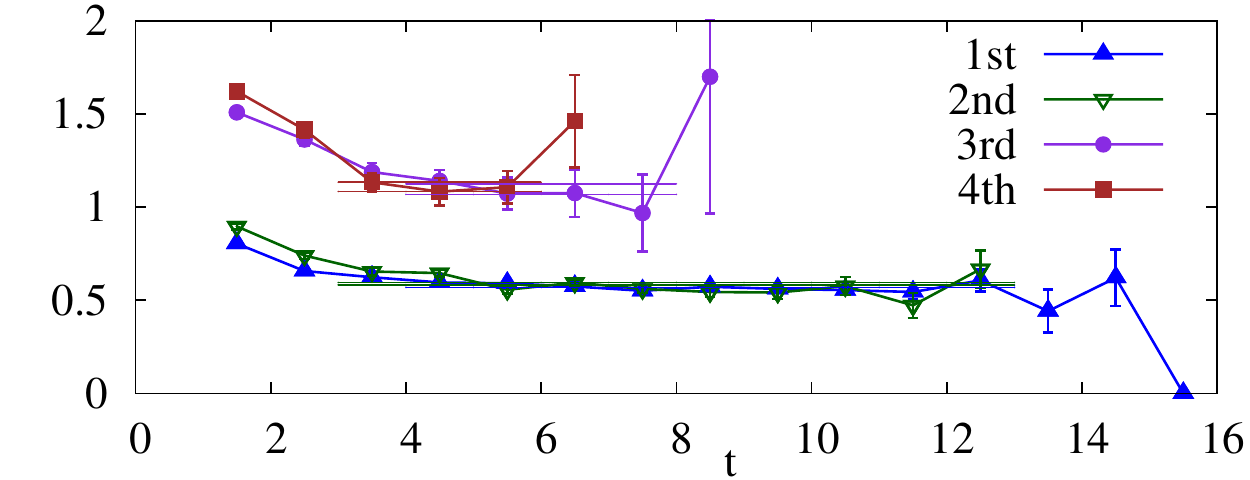}
\end{figure}

\begin{figure}
\centering  
\includegraphics[width=0.43\textwidth]{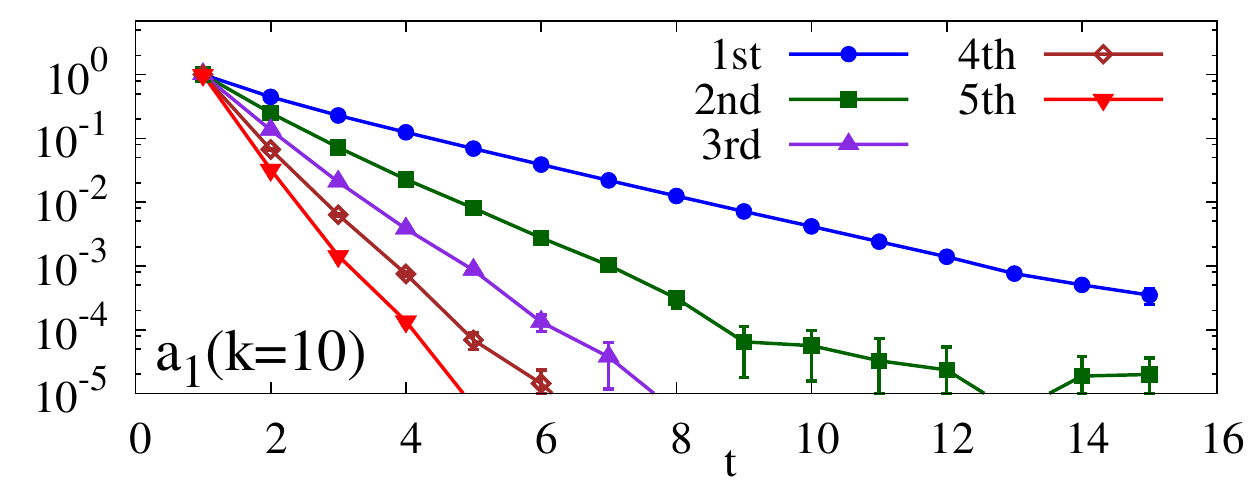}\hfill \includegraphics[width=0.43\textwidth]{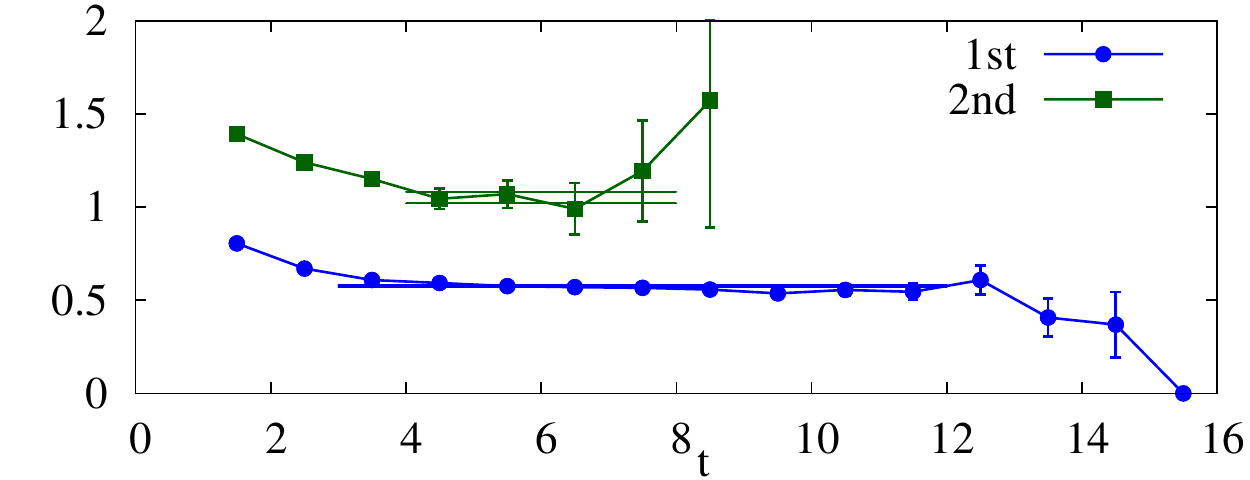}
\end{figure}
  
\begin{figure}
\centering  
\includegraphics[width=0.43\textwidth]{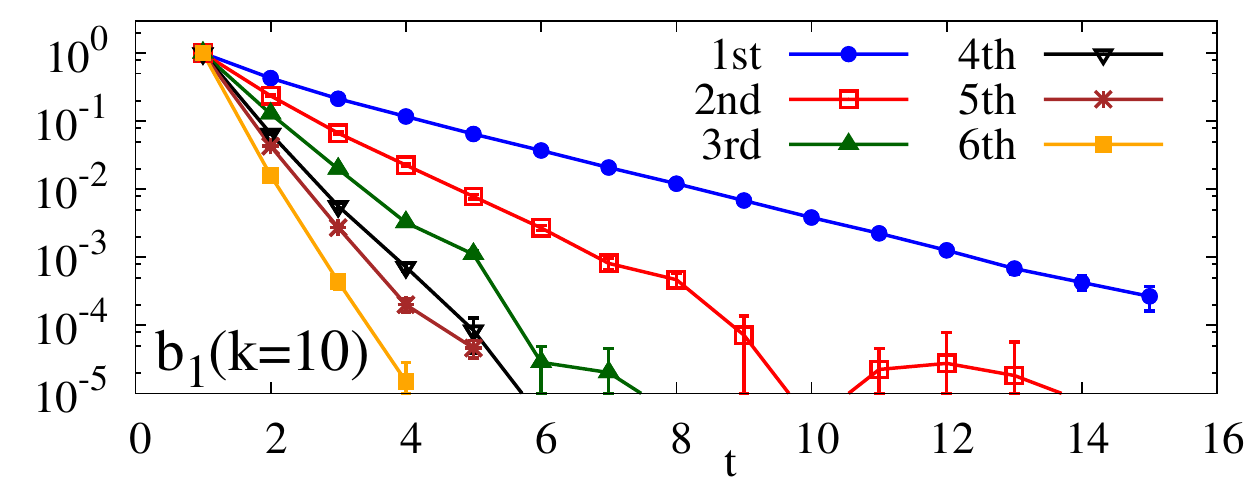}\hfill
\includegraphics[width=0.43\textwidth]{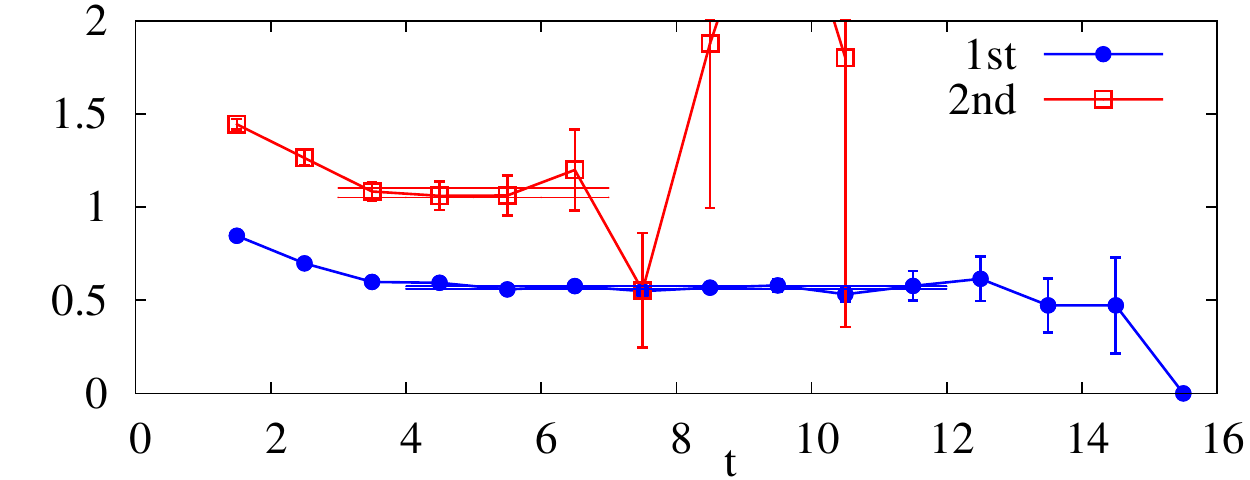}
\caption{The eigenvalues of the cross-correlation matrix and effective mass
plateaus for the isovector $J=1$ mesons with $k=10$}
\label{fig2}
\end{figure}

A very clean exponential decay of the correlators is obvious,
which means that there are physical states.
It is much cleaner than in the untruncated (real) world. The reason
for this is intuitively clear: After truncation there are no pion
fluctuations in the system. We can conclude that mesons (which are
bound states now) survive the truncation.

Note a double degeneracy of the
rho-meson eigenvalues, which is absent in the untruncated world.
This double degeneracy can be obtained only if both rho-operators
from Fig. 1 are used in the cross-correlation matrix. This double degeneracy
tells that there are two independent orthogonal degenerate $\rho$-mesons.
If we put two lowest rho-eigenvalues, the lowest $a_1$ and the lowest $b_1$
eigenvalues on the same plot, then we will see that they all are identical
(the same is true with the higher eigenvalues, but the result is less
precise). This means that there is a symmetry in the system
that is higher than  $SU(2)_L \times SU(2)_R \times U(1)_A$.

\begin{figure}
\centering 
\includegraphics[width=0.55\linewidth]{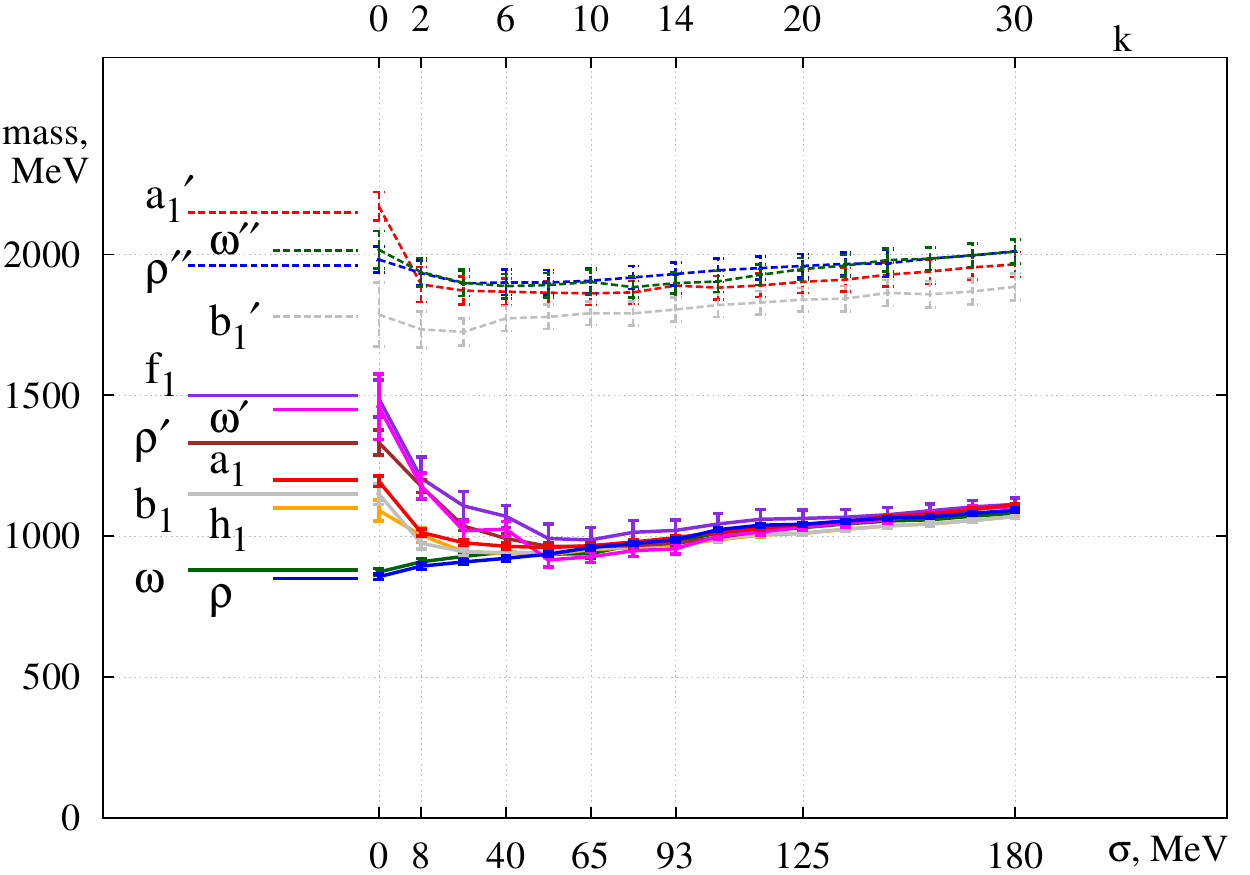}
\label{mass}
\caption{$J=1$ meson mass evolution as a function of the truncation
number $k$. $\sigma$ shows energy gap in the Dirac spectrum.} 
\end{figure}
\noindent

Evolution of meson masses is shown in Fig. 3.
We clearly see a larger degeneracy than the chiral 
$SU(2)_L \times SU(2)_R \times U(1)_A$ symmetry of the QCD
Lagrangian. What does it mean?! The same results persist for the
$J=2$ mesons and baryons.

\section{ $SU(4)$ symmetry of the meson spectra}

Given this degeneracy first we need to understand what
symmetry group does it correspond. This unexpected new symmetry
has been reconstructed in ref. \cite{g}.

Given the standard spin, parity, etc. quantum numbers we
can construct explicitly basis vectors for all irreducible
representations of the chiral group shown in Fig. 1. 

\small
{\bf (i)~~~ (0,0):}

\begin{equation}
|(0,0); \pm; J \rangle = \frac{1}{\sqrt 2} |\bar R R \pm \bar L L\rangle_J.
\end{equation}

{\bf (ii)~~~ $(1/2,1/2)_a$ and $(1/2,1/2)_b$:} 

\begin{equation}
|(1/2,1/2)_a; +;I=0; J \rangle = \frac{1}{\sqrt 2} |\bar R L + \bar L R\rangle_J,
\end{equation}

\begin{equation}
|(1/2,1/2)_a; -;I=1; J \rangle = \frac{1}{\sqrt 2} |\bar R \vec \tau L -
\bar L \vec \tau R\rangle_J,
\end{equation} 

\medskip

\begin{equation}
|(1/2,1/2)_b; -;I=0; J \rangle = \frac{1}{\sqrt 2} |\bar R L - \bar L R\rangle_J,
\end{equation}

\begin{equation}
|(1/2,1/2)_b; +;I=1; J \rangle = \frac{1}{\sqrt 2} |\bar R \vec \tau L +
\bar L \vec \tau R\rangle_J.
\end{equation}

{\bf (iii)~~~ (0,1)$\oplus$(1,0):} 

\begin{equation}
|(0,1)+(1,0); \pm; J \rangle = \frac{1}{\sqrt 2} |\bar R \vec \tau R 
\pm \bar L  \vec \tau L \rangle_J,
\end{equation}

Now we need to to find a minimal group that contains
$SU(2)_L \times SU(2)_R \times U(1)_A$ as a subgroup and 
that combines all these vectors into one irreducible
representation. These new symmetry transformations must
connect all these basis vectors. The latter requirement
can be achieved if these new symmetry transformations
mix the left- and right-handed quarks. Consequently, the
required symmetry group must contain as a subgroup the
$SU(2)_{CS}$ {\it chiralspin} rotations that act on the
following doublets:  

\begin{equation}
\textsc{U} =\begin{pmatrix} u_L \\ u_R\end{pmatrix}, ~~~~~~~
\textsc{D} =\begin{pmatrix} d_L \\ d_R \end{pmatrix}. 
\end{equation}
A three-dimensional imaginary space where these rotations are
performed is called  the {\it chiralspin}  space. The  chiralspin rotations
mix the right- and left-handed components of the fermion fields.
It is similar to the well familiar isospin space: Rotations in the isospin space mix particles with different
electric charges.

If we combine the $SU(2)_{CS}$ and the isospin $SU(2)$ group into one
larger group one arrives at the $SU(4)$ group with the fundamental
vector

\begin{equation}
\Psi =\begin{pmatrix} u_{\textsc{L}} \\ u_{\textsc{R}}  \\ d_{\textsc{L}}  \\ d_{\textsc{R}} \end{pmatrix}. 
\end{equation}

The dim=15 irreducible representation of this group connects all 
$(0,0), (1/2,1/2)_a, (1/2,1/2)_b, (1,0)+(0,1)$ vectors. One of the $(0,0)$
basis vectors, namely $|(0,0); - ; J=1 \rangle>$ is a singlet of $SU(4)$.

We can construct an explicit realization of the $SU(2)_{CS}$ and $SU(4)$
algebra that acts on Dirac spinors \cite{gp}.
Then the $SU(2)_{{CS}}$ chiralspin rotations are generated through 

$$\boldsymbol{\Sigma} = \{ \gamma^0, i \gamma^5 \gamma^0, -\gamma^5 \} \;, 
~~~~~~~~~ 
[\Sigma^i,\Sigma^j] = 2 i \epsilon^{i j k} \, \Sigma^k \; .$$

\noindent
The Dirac spinor transforms
under a global or local $SU(2)_{CS}$ transformation as
\begin{equation}
\label{V-def}
  \Psi \rightarrow  \Psi^\prime = e^{i  {\bf{\varepsilon} \cdot \bf{\Sigma}}/{2}} \Psi  \; .
\end{equation}
The $SU(4)$ group  contains at the same time $SU(2)_L \times SU(2)_R$
and $SU(2)_{CS} \supset U(1)_A$  
\noindent
and has the following set of generators:
$$ \{(\tau^a \otimes {1}_D), (1_F \otimes \Sigma^i), (\tau^a \otimes \Sigma^i) \} \; .$$
The global and local $SU(4)$ transformations of the Dirac spinor
are defined through
\begin{equation}
\label{W-def}
\Psi \rightarrow  \Psi^\prime = e^{i \bf{\epsilon} \cdot \bf{T}/2} \Psi\; .
\end{equation}

The $SU(2)_{CS}$ and $SU(4)$ transformations of all operators from Fig. 1
are shown in Fig. 4.

\begin{figure}
\centering
\includegraphics[angle=0,width=0.85\linewidth]{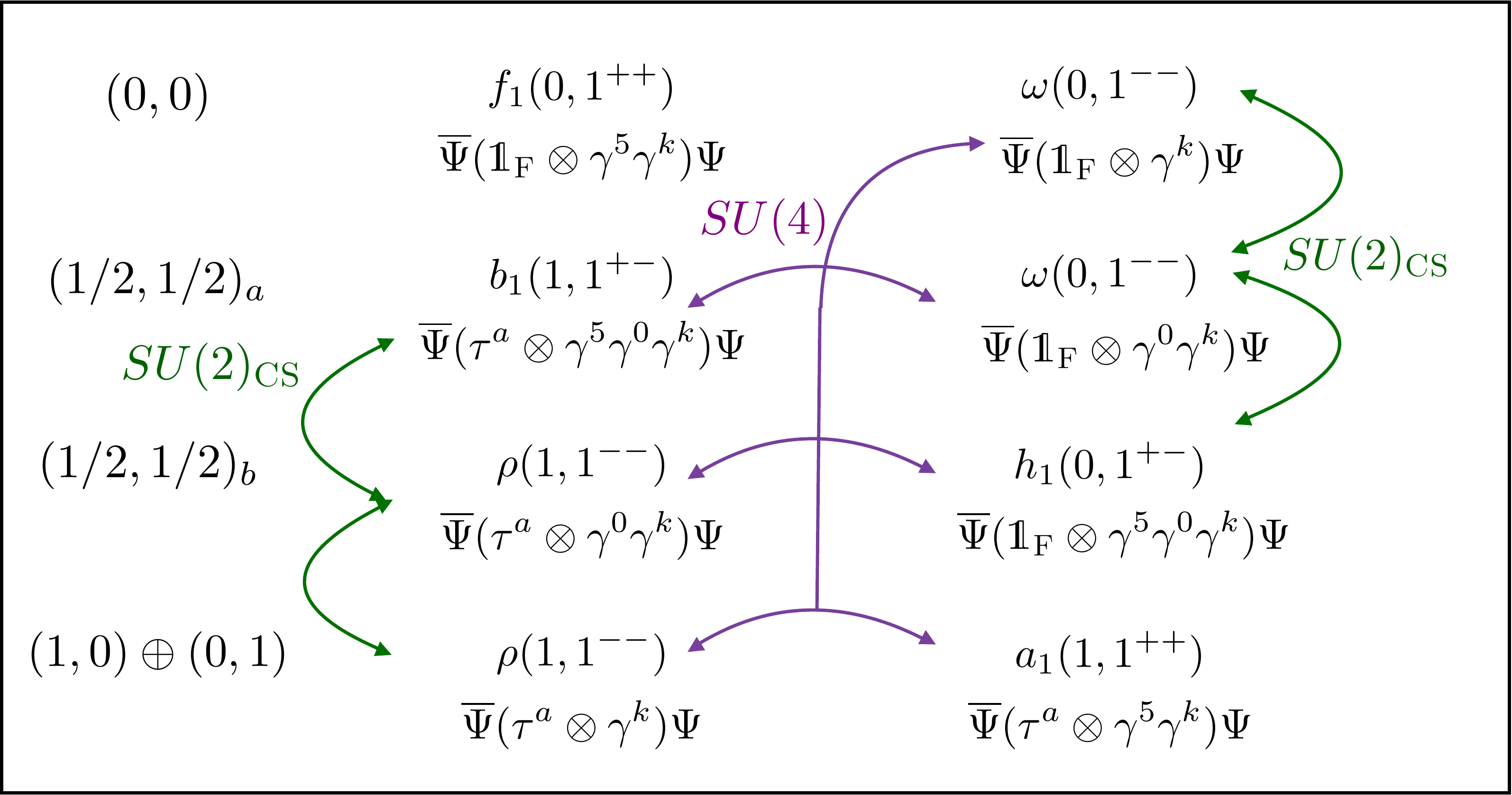}
\caption{ The green arrows connect operators that belong to the
$SU(2)_{CS}$ triplets. The $f_1$ and $a_1$ operators are the 
$SU(2)_{CS}$ singlets. The purple arrows show the $SU(4)$ 15-plet.
The $f_1$ operator is a singlet of $SU(4)$.}
\end{figure}

\section{Zero modes and hidden classical symmetries of Euclidean QCD}
The $SU(4)$ symmetry is obtained
in lattice simulations upon subtraction of the near-zero modes of the Dirac
operator. It implies that this symmetry should be encoded in the Euclidean QCD. Obviously the Lagrangian (\ref{lag}) does not have this symmetry. 
This is
because the Dirac operator does not commute with the $SU(2)_{CS}$ transformations.\footnote{In Euclidean space we have to substitute the $\gamma_0$ matrix through the $\gamma_4$ matrix.}
Then we should recover at which level this symmetry is hidden in
Euclidean QCD. More explicitly, we have to find a part of the Euclidean
QCD formalism that breaks this symmetry.

Consider the zero modes of the Dirac equation, 

\begin{equation}
 \gamma_\mu D_\mu  \Psi_0(x) = 0.
\label{dir}
\end{equation}

\noindent
Given standard antiperiodic boundary conditions for the quark
field along the time direction, the
zero modes are solutions of the Dirac equation with the gauge
configurations of a nonzero global topological charge.  
The difference of numbers of
the left-handed
and right-handed zero modes  is according to the Atiyah-Singer
theorem  fixed by the global topological charge $Q$ of the gauge configuration:

\begin{equation}
 n_L - n_R = Q.
\label{AZ}
\end{equation}

\noindent
Some $SU(2)_{CS}$ transformations rotate
the right-handed spinor into the left-handed one and vice versa.
Consequently,  the zero modes  explicitly violate the $SU(2)_{CS}$ and
$SU(2N_F)$ symmetries: 
The zero modes   introduce an asymmetry between
the left- and right-handed degrees of freedom and  break the 
$SU(2)_{CS}$ invariance. The latter  is possible only if there is no asymmetry
between the left and the right.

It is well understood, however, that the exact zero modes are completely
irrelevant since their contributions to the Green functions and
observables vanish in the thermodynamical
limit $V \rightarrow 0$ \cite{LeuS,N,A}. Consequently, in the finite
volume calculations we can subtract the irrelevant  exact zero modes.

We can expand fields $\Psi(x)$ and $\Psi^\dagger(x)$  in the Lagrangian
over a complete
and orthonormal set $\Psi_n(x)$ of the eigenvalue problem (\ref{ev}):

\begin{equation}
 \Psi(x) = \sum_{n} c_n \Psi_n(x), ~~~~~ 
 \Psi^\dagger(x) = \sum_{k}\bar {c}_k \Psi^\dag_k(x),
\end{equation}
 where $\bar {c}_k,c_n$ are Grassmannian numbers.
Then the fermionic part of the QCD partition function takes the following form

\begin{equation}
Z = \int \prod_{k,n} d\bar {c}_k dc_n  
e^{\sum_{k,n}\int d^4x 
 \bar {c}_k  c_n (\lambda_n + im) \Psi_k^\dag(x) \Psi_n(x)}. 
\label{ZZ}
\end{equation}

\noindent
Now we can directly read-off symmetry properties of the classical  part
of the partition function, i.e. of the integrand. This functional contains only a superposition of
terms  $\Psi_k^\dag(x) \Psi_n(x)$. It is precisely $SU(2)_{CS}$ and
$SU(2N_F)$ symmetric, because

\begin{equation}
(U\Psi_k(x))^\dag U \Psi_n(x) = \Psi^\dag_k(x) \Psi_n(x),
\label{tt}
\end{equation}
 
\noindent
where $U$ is any local or global transformation from the
groups $SU(2)_{CS}$ and
$SU(2N_F)$ , $ U^\dag = U^{-1}$. 
The exact zero modes, for which the equation (\ref{tt}) does not hold,
 have been subtracted from the partition function.
We conclude that classically the Euclidean QCD 
without the irrelevant exact zero mode contributions  
is invariant with respect to both
global and local $SU(2)_{CS}$ and $SU(2N_F)$ transformations.

The term  "hidden classical $SU(2N_F)$ symmetry" should be correctly understood.
It is not a global symmetry of the Lagrangian and consequently
there are no respective conserved Noether currents. However, it is a
real local symmetry of the classical part of the QCD partition function  ignoring irrelevant exact zero modes.

How is this hidden classical symmetry  broken? The integration
measure in (\ref{ZZ}) is not invariant under a local $U(1)_A$
transformation \cite{FU}. Consequently, the $U(1)_A$ anomaly
breaks the classical $U(1)_A$ symmetry.
 Since the $U(1)_A$ is a subgroup of  
$SU(2)_{CS}$, the anomaly  breaks either the $SU(2)_{CS}$ symmetry. Hence
the classical $SU(2N_F) \supset SU(N_F)_L \times SU(N_F)_R \times U(1)_A$
symmetry is broken by anomaly to $SU(N_F)_L \times SU(N_F)_R$.

The quark condensate   
 in Minkowski space
breaks all $U(1)_A$, $SU(2)_{CS}$, $SU(N_F)_L \times SU(N_F)_R$ and
$SU(2N_F)$ symmetries to the vector flavor symmetry $SU(N_F)_V$.
Hence, the new hidden classical $SU(2)_{CS}$ and $SU(2N_F)$ symmetries are
broken both by the condensate and anomalously.

Spontaneous chiral symmetry breaking is encoded in the near-zero 
modes of the Dirac operator, as it follows from the Banks-Casher
relation. If  anomaly is also encoded in
the near-zero modes, as suggested e.g. by the instanton mechanism of
both breakings, then
removal on lattice of the near-zero modes
should restore not only chiral $SU(N_F)_L \times SU(N_F)_R$  and $U(1)_A$ symmetries, but also
a larger $SU(2N_F)$ symmetry, which naturally explains lattice observations reviewed in previous sections.

\section{What symmetries should one expect in mesons and baryons
upon truncation of the near-zero modes?}

From the results presented in Fig. 3 it is clearly seen that the
degeneracy pattern is larger than $SU(4)$, because the $SU(4)$ singlet
($f_1$) and the $SU(4)$ 15-plet mesons ($\rho,\rho',a_1,b_1,h_1,\omega,\omega'$)
are also degenerate. This implies that actually some higher symmetry
is observed that includes the $SU(4)$ as a subgroup \cite{gp}. It was found
that no higher symmetry  exists,  that
would connect {\it local} quark bilinears from the 15-plet and singlet of $SU(4)$
within  the same irreducible representation \cite{TDC}. 

This challenging problem has been solved in ref. \cite{gg}.
Hadron spectra are extracted from the correlation functions calculated
with the gauge-invariant source operators. At each time slice "t" a meson
correlator contains minimum the lowest Fock  $\bar q q$ component 
with a quark and an antiquark  located at different space points 
$\bf{x}$ and $\bf{y}$. Both $q$ and $\bar q$ interact with the same gauge configuration. Then all
arguments of the previous section apply independently for $q$ and $\bar q$. Since the $SU(2N_F)$ invariance is local, we can perform  $SU(2N_F)$ rotations at points $\bf{x}$ and $\bf{y}$ with different rotation parameters. It is then clear that
the meson correlation function with the $\bar q q$ valence content
has a {\it bilocal} $SU(2N_F) \times SU(2N_F)$ symmetry. A symmetry of 
higher Fock components is obviously larger, but the whole correlator 
has a symmetry of the lowest $\bar q q$ component. Obviously,
averaging over gauge configurations does not change  this symmetry property.

The same argument applies to baryons and in this case we  expect a trilocal
$SU(2N_F) \times SU(2N_F) \times SU(2N_F)$ symmetry. 

One of the
irreducible representations of the $SU(4) \times SU(4)$ is 16-dimensional
and is a direct sum of the 15-plet and singlet of $SU(4)$. Hence a direct prediction of this bilocal  symmetry is a degeneracy of the $SU(4)$-singlet and
of the $SU(4)$ 15-plet, in agreement with the lattice observations. 
This symmetry is bilocal and cannot be represented
by the local composite operators which is consistent with  
conclusions of Ref. \cite{TDC}. 

\section{A short summary of our findings.}

Our main findings can be summarized as follows.

1. The classical part of the  partition function (the integrand), excluding  irrelevant exact zero mode
contributions, has  $SU(2)_{CS}$ and $SU(2N_F)$ local symmetries. Since these symmetries
are not symmetries of the QCD Lagrangian  we refer them  as  hidden classical symmetries of QCD. There are no respective conserved Noether currents.
These symmetries are broken at the quantum level
by the axial anomaly  and by the quark condensate.
The physics of chiral symmetry spontaneous breaking and of anomaly is
contained in the  near-zero modes of the Dirac operator. Their truncation
on the lattice should restore not only the 
$SU(N_f)_L \times SU(N_F)_R \times U(1)_A$ chiral symmetry 
but actually higher hidden classical symmetries $SU(2)_{CS}$ and $SU(2N_F)$.

2. We have shown
that elimination of the near-zero modes  leads
to  $SU(2N_F) \times SU(2N_F)$ and 
$SU(2N_F) \times SU(2N_F) \times SU(2N_F)$
symmetries in mesons and baryons.

3. The bilocal $SU(4) \times SU(4)$ symmetry  explains a
degeneracy of the $SU(4)$ singlet $f_1$ correlator with the $SU(4)$
15-plet $\rho, \rho', \omega, \omega', h_1, a_1, b_1$  correlators.

\section{Implications}

It is natural to expect many different implications of the
hidden classical symmetry. Here we will mention a most dramatic one \cite{gg2}. 

At high temperature the  quark condensate of the vacuum vanishes. 
There are lattice indications that above the critical temperature
the $U(1)_A$ symmetry is restored and a gap opens in the Dirac spectrum \cite{A1,A2}.\footnote{See also the
opposite statement \cite{K} and its critique \cite{A4}.}
Then
it follows that the $SU(2)_{CS}$ and $SU(2N_F)$
symmetries are manifest in Euclidean correlation functions and observables.
Such symmetries cannot be obtained in terms of deconfined quarks and gluons
in Minkowski space,  where we live. Hence at high temperatures QCD is also
in the confining regime and elementary objects are color singlet
$SU(4)$ symmetric "hadrons". "Hadrons" with such a symmetry can be directly
constructed in Minkowski space \cite{sh}.

\bigskip
We acknowledge a partial support from the Austrian Science Fund (FWF)
through the grant P26627-N27.

\end{document}